\documentclass[aps,pra,superscriptaddress,twocolumn,showpacs]{revtex4}

\renewcommand*{\[}{\begin{equation}}

\renewcommand*{\]}{\end{equation}}

\newcommand{\myscaleboxb}[1]{\scalebox{0.7}[0.7]{#1}}

\usepackage{epsfig}

\begin{document}

\title{Influence of gas pressure on high-order harmonic generation of Ar and Ne}

\author{Guoli Wang}
\affiliation{J. R. Macdonald Laboratory, Physics Department, Kansas
State University, Manhattan, Kansas 66506-2604, USA}
\affiliation{Key Laboratory of Atomic and Molecular Physics and
Functional Materials of Gansu Province, College of Physics and
Electronic Engineering, Northwest Normal University, Lanzhou,730070,
China}

\author{Cheng Jin}
\affiliation{J. R. Macdonald Laboratory, Physics Department, Kansas
State University, Manhattan, Kansas 66506-2604, USA}

\author{Anh-Thu Le}
\affiliation{J. R. Macdonald Laboratory, Physics Department, Kansas State University, Manhattan, Kansas 66506-2604, USA}

\author{C. D. Lin}
\affiliation{J. R. Macdonald Laboratory, Physics Department, Kansas State University, Manhattan, Kansas 66506-2604, USA}

\date{\today}

\begin{abstract}
 We study the effect of gas pressure on the generation of high-order
 harmonics where harmonics due to individual atoms are calculated
 using the recently developed quantitative rescattering theory, and
 the propagation of the laser and harmonics in the  medium is calculated
 by solving the Maxwell's wave equation. We illustrate that the simulated
 spectra are very sensitive to the laser focusing conditions at high
 laser intensity and high pressure since the fundamental laser
 field is severely reshaped during the propagation. By comparing the simulated
 results with several experiments we show that the pressure
 dependence can be qualitatively explained. The lack of quantitative
 agreement is tentatively attributed to the failure of the complete knowledge of the
 experimental conditions.
\end{abstract}


\maketitle

\section{Introduction}

In the last two decades, high-order harmonic generation (HHG) has
been widely studied for its potential as a short-wavelength tabletop
light source~\cite{Tenico,Tenico2,light-source}, or as an ultrashort
attosecond pulse or pulse train \cite{Atto1,Atto2}. The harmonic
emission in a gas medium is well understood qualitatively. Harmonics
are emitted when atoms or molecules are exposed to an intense
infrared laser field. These lights travel collinearly with the input
laser and interact with the generating medium. Thus a full
description of HHG involves a quantum treatment of the harmonic
emission from a single atom or molecule, together with the
propagation of the harmonics and the nonlinear interaction of the
laser light with the medium. Recently, Jin {\it et al.}
\cite{Jin09,JinJPB,JinPRA2011,Jin-CO2} have combined the
well-established propagation theory with the quantitative
rescattering (QRS) theory \cite{qrs1,qrs2,qrs3} for individual atoms
and molecules, to provide a quantitative description of HHG
generated from atomic and molecular targets by an intense laser
pulse. Its success has been shown by comparing with the recent
experimental measurements \cite{JinJPB,JinPRA2011,Jin-CO2}. These
applications have focused on the region where the fundamental laser
field is not severely modified in the medium. Such studies are
useful for using HHG to probe the structure of molecules. For
applications, it is desirable to achieve highest number of photons
by increasing the laser intensity and the gas pressure. If the HHG
from individual atoms are fully phase-matched, then the number of
photons will increase quadratically with the medium pressure.
Unfortunately, the nonlinear interaction of lasers with the gas
medium is complicated. The intense laser light and the harmonics can
be dispersed and absorbed by the medium. The intense laser light
also ionizes the atoms (or molecules) and generates free electrons,
thus changing the optical properties of the gas medium. These
nonlinear interactions of the laser light and the harmonics with the
medium can be calculated by solving the proper Maxwell's wave
equations, with the induced dipole from each atom or molecule by the
laser as the source of the harmonics. A detailed theory of harmonic
generation with the inclusion of macroscopic propagation where
single-atom HHG emissions are calculated with the QRS theory has
been given in Ref. \cite{JinPRA2011}.

In this paper our goal is to extend this theory to the harmonic
generation for laser intensities near and above the critical
intensity for a given target. In particular, we want to investigate
the effect of pressure on the yield of HHG, as the intensity is
varied or when the focusing conditions are changed. At high
intensities, HHG is a highly nonlinear process, and the harmonic
yields are very sensitive to all the parameters of the experiment.
These parameters, unfortunately, are not always well specified in an
experiment. We perform simulations by varying some parameters in
order to achieve optimal agreement with the reported experimental
data. We also check the effects of various parameters of the medium
to illustrate how these individual parameters change the simulated
HHG spectra. These kinds of studies have been carried out
previously, sometimes connecting with experimental observations.
However, in these calculations HHG from each atom was often
calculated using the strong field approximation (SFA), or the
so-called Lewenstein model \cite{SFA}. It is well known that the SFA
does not predict correct HHG spectra from individual atoms. By using
QRS, the atomic dipoles induced by the laser are accurately
calculated. These induced dipoles are then fed into the Maxwell's
propagation equations. We perform simulations to check to what
extent some of the earlier experimental HHG spectra can be simulated
with the present theory.

In Section~\ref{Theory}, we briefly summarize how the calculations
are done. We then simulate the HHG spectra of Ar for 6-fs pulses at
quite high intensity reported in Ref. \cite{Altucci05}. We also
examine the effect of gas pressure on the behavior of Cooper minimum
at high intensities. For higher pressure the laser field is severely
modified as it propagates through the medium. We show that the
Cooper minimum eventually disappears at pressure above about 50
Torr. The disappearance of the Cooper minimum is entirely due to the
medium propagation effect \cite{Japan}. Next we examine the pressure
effect on the harmonics, and it is established in general that there
is an optimum pressure for each harmonic beyond which the harmonic
yield drops with increasing pressure. Since the propagation of
harmonics is controlled by the optical properties of the medium, we
check if there is a single factor that would mostly affect the
harmonic generation. We also take a close look at the dispersion and
absorption coefficients used in the simulation. These tabulated
data, which were obtained from theoretical calculations, have not
been carefully calibrated and should be used with caution. Finally
we examine the propagation of HHG generated in Ne at high intensity
and high gas pressure to compare with the result reported in a
recent experiment \cite{Dachraoui09}. We have found that in this
case HHG spectra are very sensitive to the experimental parameters
such that simulation can only achieve  qualitative agreement.

\section{Theoretical method}\label{Theory}
The detailed description of the theoretical method used in the
calculation can be found in Ref. \cite{JinPRA2011}. Briefly, we
consider both the propagation of the fundamental and the harmonic
fields in an ionizing medium. For the fundamental field, we include
dispersion, absorption, Kerr, and plasma effects. For the harmonics,
the dispersion and absorption due to neutral atoms are included. The
induced dipoles for single atoms are obtained using the QRS theory
\cite{qrs2}:
\begin{eqnarray}
D^{\rm{QRS}}(\omega)=W^{\rm{SFA}}(\omega) d^{\rm{QRS}}(\omega).
\end{eqnarray}
Here $W^{\rm{SFA}}(\omega)$ is the wave packet calculated from the
Lewenstein model \cite{SFA}, normalized by the tunneling ionization
rate calculated using the Ammosov-Delone-Krainov (ADK) theory
\cite{ADK,Tong}. The transition dipole moment $d^{\rm{QRS}}(\omega)$
is obtained by treating the real atom in a single-active electron
approximation. For the argon target used in the present work,
$d^{\rm{QRS}}(\omega)$ is calculated by using the model potential
given by M\"{u}ller \cite{Muller}. For Ne, the model potential takes
the form given in Refs. \cite{Tong,Jin09}. The resulting induced
dipoles for hundreds of different peak intensities are then fed into
Maxwell's wave equations. We assume that the laser beam at the
entrance of the gas jet has the Gaussian shape both in time and
space. The harmonics emitted at the exit face of the gas jet may
propagate further in free space until they are detected. Thus the
harmonic spectra observed experimentally also depend on the
detecting conditions \cite{JinPRA2011}. In this paper, except for
one case to be described below, all the harmonic signals (i.e., the
total signal) from the exit of the medium are collected.

\section[]{Results and discussions}

\subsection{Macroscopic HHG spectra of Ar: Theory vs experiment}

\begin{figure}
\mbox{\rotatebox{0}{\myscaleboxb{
\includegraphics[width=0.60\textwidth]{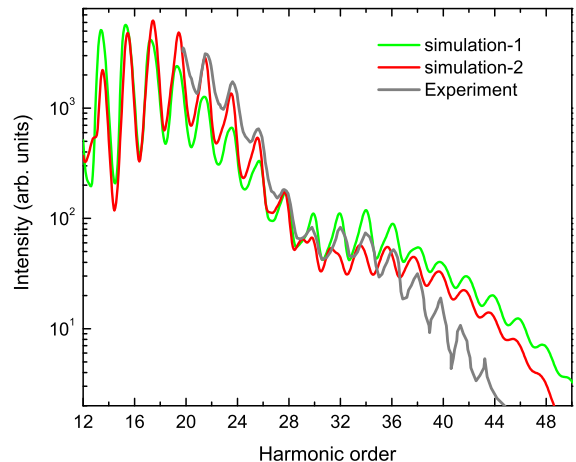}}}} \vskip-0mm
\caption{Comparison of theoretical and experimental HHG spectra of
Ar generated by a 760-nm and 6-fs laser. Experimental data are from
Ref. \cite{Altucci05}. For simulation-1 and simulation-2 the laser
peak intensity at focus is 4.5 and 4.0$\times$10$^{14}$ W$/$cm$^2$,
the length of gas jet is 1 and 2 mm, respectively. The other laser
parameters are given in the text.} \label{fig1}
\end{figure}

We first simulate experimental HHG spectra of Ar generated by a
760-nm and 6-fs (FWHM) laser pulse reported in \cite{Altucci05}.
Experimentally, the gas jet was placed 2 mm after the laser focus.
The confocal parameter of the laser beam was 10 mm (the beam waist
was 34.8 $\mu$m if a gaussian beam is assumed). The laser intensity
at the focus was estimated to be 6$\times$10$^{14}$ W$/$cm$^2$. In
simulation-1, we use the gas-jet width of 1 mm, and the gas pressure
is 40 Torr. We find that a laser peak intensity of
4.5$\times$10$^{14}$ W$/$cm$^2$ would give better agreement with the
spectra in the cutoff region. The calculations have been carried out
for different carrier-envelope phases (CEPs) since a short 6-fs
laser pulse was applied in the experiment. The theoretical spectra
are finally averaged over all the CEPs. To compare with experiment,
the HHG spectra are normalized at harmonic 27 (H27). We can see that
the main features in the experimental spectra are well reproduced by
simulation-1. However, the overall slope of the harmonic yields is
not the same. The simulated harmonic yields are too low for lower
orders, but too high for higher orders. But this result is already a
great improvement over the theoretical simulation reported in Fig.
4(c) of Ref. \cite{Altucci05}. Since HHG spectra are sensitive to
experimental conditions, we thus make another simulation where the
gas-jet length is extended to 2 mm and the laser intensity is
reduced to 4.0$\times$10$^{14}$ W$/$cm$^2$. In this simulation-2,
other parameters are the same as those in simulation-1. We can see
that simulation-2 gives better agreement for the low harmonics up to
H27, but the slope for higher harmonics is still not reproduced
correctly. We have made further adjustment of the laser parameters
within the capability of our present code but were unable to improve
agreement. At present our code is limited to a Gaussian (spatial)
input pulse. In the experiment of Altucci {\it et al.}
\cite{Altucci05}, the laser was coupled into an argon-filled
capillary consisting of a 60-cm-long fiber whose spatial profile
would be the Bessel function. Thus the emerged 6-fs pulse may be
closer to a truncated Bessel beam. It remains to be seen if the
discrepancy can be removed if a truncated Bessel beam is assumed for
the input pulse~ \cite{Bessel1,Bessel2,Bessel3}.

Below we aim at investigating the gas pressure effects on the HHG
spectra.

\subsection[]{Influence of the gas pressure on Cooper minimum and spectral modulation}
\begin{figure*}
\mbox{\rotatebox{0}{\myscaleboxb{
\includegraphics[width=1.27\textwidth]{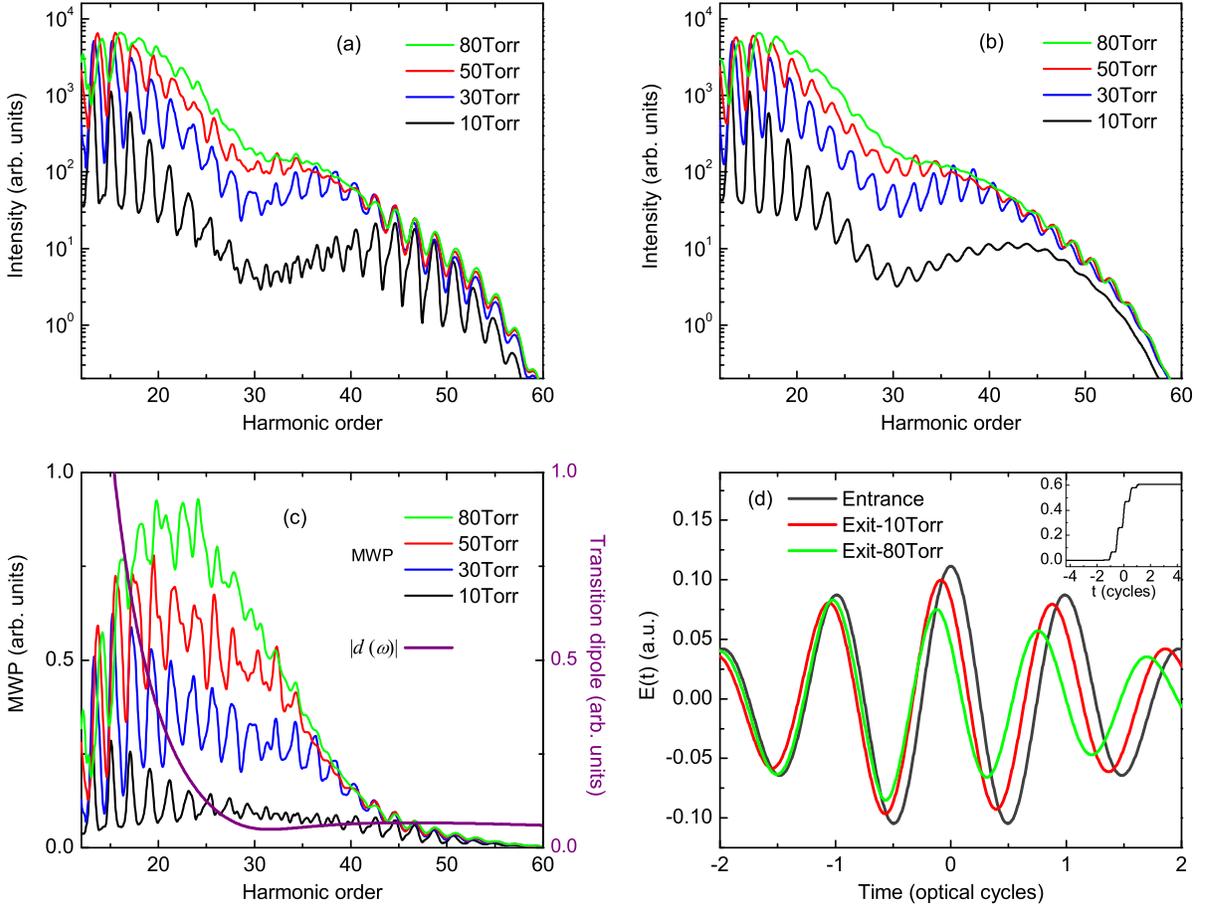}}}} \vskip-0mm
\caption{Macroscopic HHG spectra of Ar at the gas pressures of 10,
30, 50, and 80 Torr for (a) CEP=0 and (b) CEP averaged. (c) The
normalized macroscopic wave packet (MWP) at four different pressures
for CEP=0 and transition diploe moment $|d(\omega)|$. (d) Evolution
of the on-axis electric field: at entrance and exit. The inset shows
the ionization probability vs time at entrance. The laser peak
intensity at focus is 5$\times$10$^{14}$W/cm$^2$. The other
parameters are the same as those in simulation-1 in
Fig.~\ref{fig1}.} \label{fig2}
\end{figure*}

In Figs.~\ref{fig2}(a) and \ref{fig2}(b) we show the macroscopic HHG
spectra of Ar for four different gas pressures: 10, 30, 50, and 80
Torr, at a laser intensity of 5$\times$10$^{14}$ W$/$cm$^2$ at focus
for CEP=0 and CEP averaged, respectively. The other parameters are
the same as those in simulation-1 of Fig.~\ref{fig1}. Since a
few-cycle pulse is used, the harmonic spectra show a strong CEP
dependence in the cutoff region. After CEP average the spectra
become very smooth. There are two additional features in
Figs.~\ref{fig2}(a) and \ref{fig2}(b) that warrant further
discussions.

The first is about the well-known Cooper minimum (CM) of the HHG in
Ar. Although CM in harmonic spectra is a signature of the electronic
structure of the Ar atom \cite{Hans-prl-09,Colosimo-NatPhys08},
previous experimental   studies have shown that the ``visibility" of
CM is sensitive to experimental conditions. For example, the
position of the gas jet \cite{stanford} and the gas pressure
\cite{Japan} could wash out the CM seen in the harmonic spectra. It
has also been pointed out  by Jin {\it et al.} \cite{JinPRA2011}
that the CM also depends on how the HHG spectra are measured, i.e.,
in the near or the far fields, with or without the slit. The present
simulations show a similar CM dependence on gas pressure, which has
been observed in Ref. \cite{Japan}, although the experiment was
performed under different   conditions (laser duration was $\sim$50
fs, and the laser intensity was $\sim$2.8$\times$10$^{14}$
W$/$cm$^2$). In Figs.~\ref{fig2}(a) and \ref{fig2}(b), there appears
an obvious minimum around 31st harmonic ($\sim$51 eV) below 30 Torr.
When the pressure is increased to 50 Torr, the minimum becomes
barely visible, and then it is totally washed out when the pressure
is increased to 80 Torr. According to the QRS
\cite{JinJPB,JinPRA2011}, these variations are attributed to the
change of the macroscopic wave packet (MWP). The QRS states that the
macroscopic HHG spectrum can be expressed as
\cite{Jin09,JinJPB,JinPRA2011}
\begin{eqnarray}
S_{\rm{h}}(\omega)\propto \omega^4|W(\omega)|^2|d(\omega)|^2,
\end{eqnarray}
where $W(\omega)$ is the MWP, which reflects the effect of laser and
experimental conditions, and $d(\omega)$ is the photorecombination
(PR) transition dipole moment, which is a property of the target
only. In Fig.~\ref{fig2}(c) we show the detailed comparison of the
MWPs at four gas pressures for CEP=0. It shows that the MWP
($|W(\omega)|$) varies with the gas pressure. Around photon energy
of 50 eV ($\sim$H31), the MWP is very flat at 10 Torr. With
increasing gas pressure it becomes more steeper. Meanwhile, we also
show the PR transition dipole moment $|d(\omega)|$ for reference. It
is concluded that the rapid change of the MWP around 50 eV causes
the disappearance of the CM in the HHG spectra.

The second feature in Fig.~\ref{fig2} is the spectral modulation. In
Fig.~\ref{fig2}(a), for harmonics above the CM, the spectra
modulation is large at 10 Torr. The spectra are less modulated at 30
and 50 Torr, and become almost continuous when the gas pressure is
increased to 80 Torr. These continuous HHG spectra could be used to
produce isolated attosecond pulses (IAPs) with a CEP-stabilized
laser. Note that the continuum structure in HHG spectra of Xe has
been discussed by Jin {\it et al.} \cite{Jin-Atto} recently [see
Fig. 1 in that paper]. They found that  reshaping (blue shift and
defocusing) of the fundamental laser field was responsible for these
phenomena, and they also showed a method to produce an IAP by
spatial filtering in the far field. To check this mechanism in our
situation, we show the evolution of the on-axis electric field at
the entrance and exit of the gas jet for 10 Torr and 80 Torr in
Fig.~\ref{fig2}(d). Note that the time is defined in the moving
coordinate frame~\cite{JinPRA2011,Jin-Atto}. The ionization
probability is very high (about 61$\%$ at the end of a laser pulse
with peak intensity of 4.63$\times$10$^{14}$ W$/$cm$^2$). The
electric field has a good Gaussian form at the entrance. In the
leading edge of laser pulse, where the ionization probability is
very small as seen in the inset of Fig.~\ref{fig2}(d), the electric
field at the exit face has a small shift with respect to the one at
the entrance. This shift is originally from the geometric phase due
to the tightly focused laser beam. In the falling edge, there is an
obvious blue shift, and the electric field at 80 Torr is much
reduced than that at 10 Torr. We can conclude that the fundamental
laser field is reshaped. Combined with effects due to CEP average,
our simulations in Fig.~\ref{fig2}(b) show that spectral modulation
above the CM increases from 10 Torr to 30 Torr, and then decreases
with increasing gas pressure until the HHG spectra become continuous
at 80 Torr. Similar phenomena have been observed in Zheng {\it et
al.} \cite{APL} for a phase-stabilized few-cycle laser pulse.

\subsection[]{Gas pressure dependence of HHG conversion efficiency at fixed laser intensity and gas-jet length}

\begin{figure*}
\includegraphics[width=0.85\textwidth]{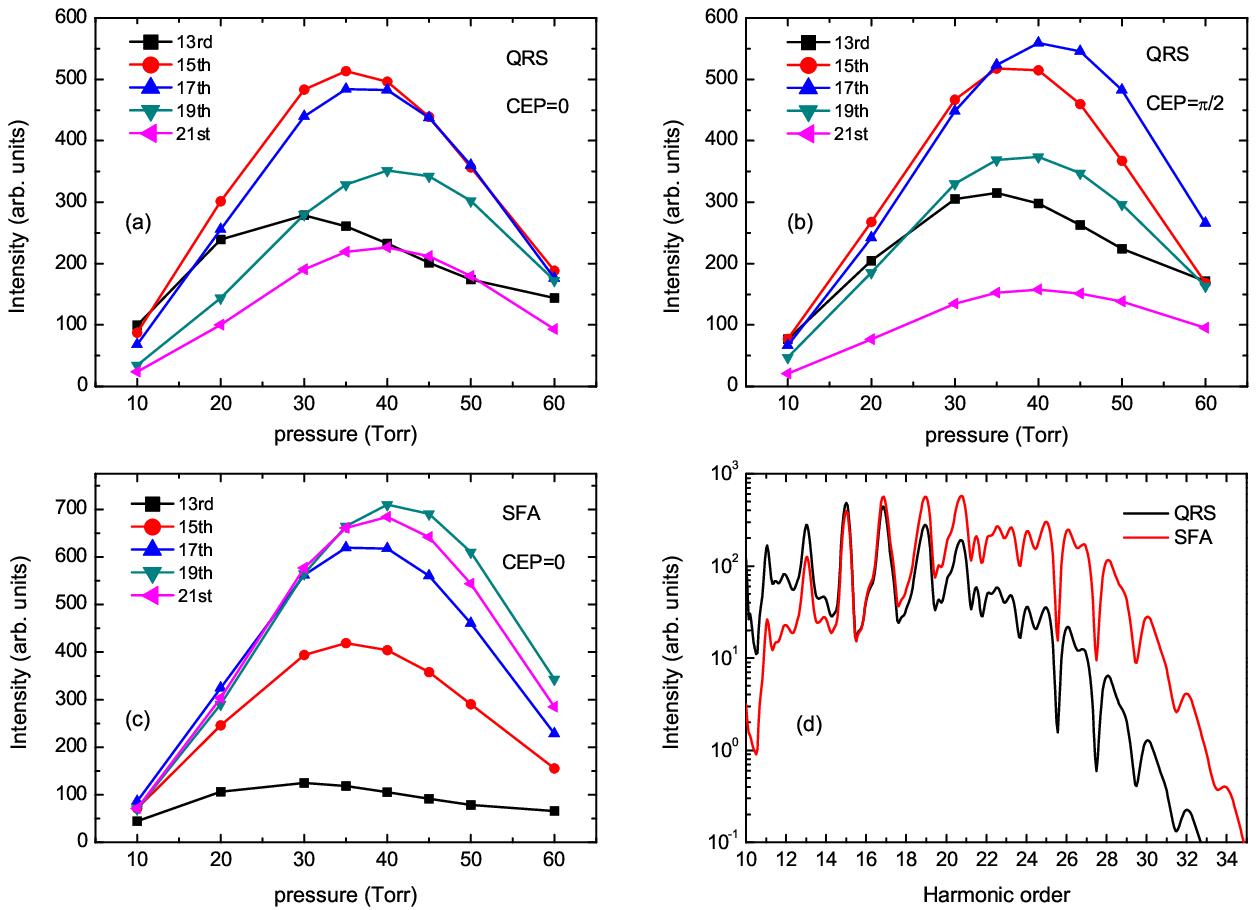} \vskip-0mm
\caption{The harmonic intensity of Ar as a function of gas pressure
for different orders from 13rd to 21st in the plateau for (a) CEP=0
and (b) CEP=0.5$\pi$ obtained by QRS. The laser peak intensity at
focus is 2$\times$10$^{14}$ W$/$cm$^2$, the other parameters are the
same as those in simulation-1 in Fig.~\ref{fig1}. (c) The same as
(a) but obtained by SFA. (d) Comparison of HHG spectra obtained by
QRS and SFA for a pressure of 30 Torr and CEP=0.} \label{fig3}
\end{figure*}

\subsubsection{The optimal HHG yield vs gas pressure}
The macroscopic harmonic generation is strongly influenced by many
factors collectively. In other words, the conversion efficiency
cannot be improved by just increasing the laser intensity, atomic
density, and medium length independently. There are three physical
effects that limit harmonic generation: absorption, defocusing, and
dephasing \cite{Krausz2000}. Many experimental and theoretical
studies that investigated harmonics efficiency (or photon flux) have
been carried out, such as absorption by Schn\"{u}rer {\it et al.}
\cite{Schnurer99,Schnurer2000}, defocusing by Altucci {\it et al.}
\cite{Altucci96} and Dachraoui {\it et al.} \cite{Dachraoui09}, and
dephasing in a hollow waveguide in Refs.
\cite{Rundquist98,Durfee99}.

Figure~\ref{fig3}(a) shows the dependence of harmonic intensity on
the gas pressure for different plateau harmonic orders between H13
and H21 for CEP=0 at the intensity of 2$\times$10$^{14}$ W$/$cm$^2$.
The gas jet is 1-mm long, with other parameters the same as those in
simulation-1 of Fig.~\ref{fig1}. With the increase of gas pressure,
the yield for each harmonic order increases first to reach a maximum
and finally drops with even higher pressure. The maximum yield
shifts to a higher pressure for higher harmonic order, especially in
the region of low harmonic orders. These have been demonstrated
experimentally in an Ar gas-filled hollow waveguide
\cite{Rundquist98} and in a gas cell \cite{Xu04}. We show a similar
figure for CEP=$\pi$/2 in Fig.~\ref{fig3}(b).

The above results are most easily understood from a one-dimensional
model given by Constant {\it et al.} \cite{Constant}. For the $q$th
harmonic order, the number of photons is proportional to [Eq. 1 of
Ref. \cite{Constant}]
\begin{eqnarray}
\label{con-model}&\rho^{2}A_{q}^{2}&\frac{4L_{\rm{abs}}^{2}}{1+4\pi^{2}(L_{\rm{abs}}^{2}/L_{\rm{coh}}^{2})}\bigg[1+{\rm{exp}}\Big(-\frac{L_{\rm{med}}}{L_{\rm{abs}}}\Big)
\nonumber\\&&-2{\rm{cos}}\Big(\frac{\pi
L_{\rm{med}}}{L_{\rm{coh}}}\Big){\rm{exp}}\Big(-\frac{L_{\rm{med}}}{2L_{\rm{abs}}}\Big)\bigg],
\end{eqnarray}
where $A_{q}(z)$ (units: C$\cdot$m) is the amplitude of the atomic
response. Here \emph{L}$_{\rm{coh}}$=$\pi/\Delta$$k_q$ is the
coherence length, where $\Delta$$k_q$ is the wave vector mismatch
between the fundamental and the generated harmonic field,
\emph{L}$_{\rm{abs}}$=1$/\sigma\rho$ is the absorption length, where
$\sigma$ and $\rho$ are the photoionization cross section and the
density of the generating gas, respectively. According to this
model, if we assume $A_{q}(z)$ to be independent of $z$ (this is
true for a loosely focused laser beam), then the harmonic yield is
determined by the three parameters: the medium length
\emph{L}$_{\rm{med}}$, coherence length \emph{L}$_{\rm{coh}}$, and
absorption length \emph{L}$_{\rm{abs}}$. The phase mismatch
$\Delta$$k_q$ is determined by different dispersion terms: atomic,
electronic, geometric dispersion (the so-called Gouy phase for a
Gaussian beam), and by the gradient of the atomic dipole phase
\cite{Dachraoui09}. In Fig.~\ref{fig3}, the medium length is fixed.
An increase of pressure makes the absorption length smaller.
According to Fig.~1 of \cite{Constant}, the harmonic yield would
reach saturation, the larger the phase mismatch, the smaller the
number of photons generated for the harmonic. The results from
Fig.~\ref{fig3} which are simulated from the 3D model, show that as
the pressure increases beyond the optimal pressure for the
generation of a given harmonic, continuing increase of pressure
makes the harmonic yield drop. In fact, since the coherent length
varies spatially for each harmonic, the spatial distribution of the
harmonic signal also changes with  pressure. In fact many factors
will influence coherence length or absorption length through their
dependence on gas pressure. Furthermore, the coherence length and
absorption length also depend on harmonic order. All of these can
affect the change of optimal pressure with harmonic order. If the
single-atom harmonics are calculated using SFA, the pressure
dependence of each single harmonic will be qualitatively correct,
see Fig.~\ref{fig3}(c). On the other hand, the relative intensities
between the harmonics would not be correct since SFA does not use
correct photoabsorption cross sections. Thus the HHG spectra
predicted using SFA and QRS are different, see Fig.~\ref{fig3}(d).

\subsubsection{Dependence of harmonic spectra on the optical properties of the gas medium}

\begin{figure}
\includegraphics[width=8.5cm]{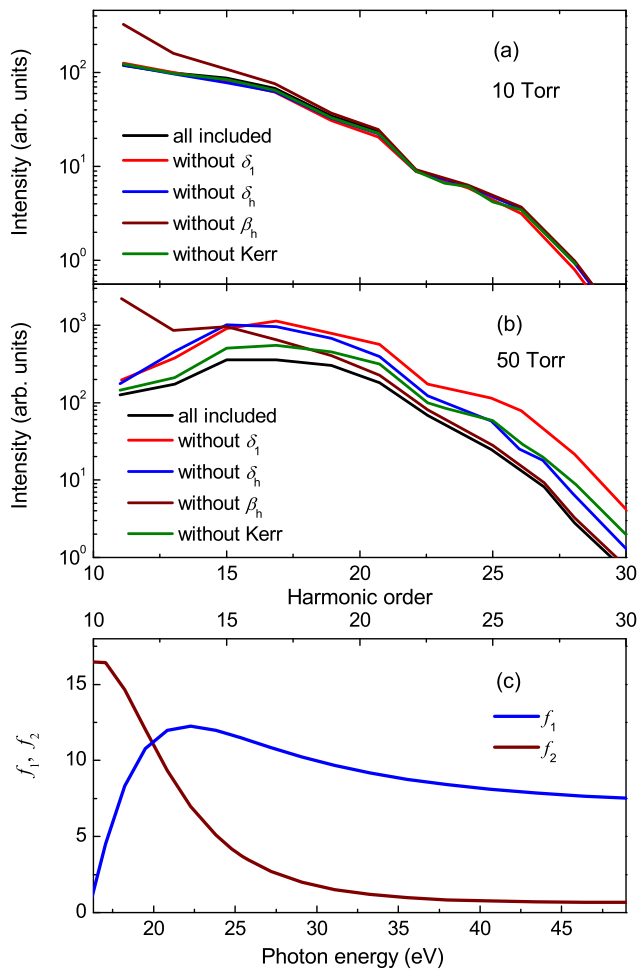} \vskip-0mm
\caption{The influence of dispersion of the fundamental ($\delta_1$)
and harmonic ($\delta_{\rm h}$) field, absorption of harmonics
($\beta_{\rm h}$), and Kerr effect at (a) 10 Torr and (b) 50 Torr.
The parameters are the same as Fig.~\ref{fig3}(a). (c) The atomic
scattering factors $f_1$ and $f_2$ are from NIST~\cite{NIST} for
Ar.} \label{fig4}
\end{figure}

In the calculations above, few-cycle laser pulses are applied and
laser intensity is low such that free electron density in the gas
medium is very small. We have checked that effects due to free
electrons (defocusing and phase mismatch) in these calculations are
negligible. In the simulation we include the dispersion
($\delta_{1}$), Kerr nonlinearity ($\eta_{2}$) of the fundamental
field, the dispersion ($\delta_{\rm{h}}$) and absorption
($\beta_{\rm{h}}$) of the harmonic fields in the simulation
\cite{JinPRA2011}. All of these factors are pressure dependent. Note
that $\delta_{1}$, $\eta_{2}$, and $\delta_{\rm{h}}$ all contribute
to phase mismatch $\Delta$$k_q$; in other words, they all affect the
coherence length \emph{L}$_{\rm{coh}}$. In the meanwhile,
$\beta_{\rm{h}}$ leads to pressure-dependent \emph{L}$_{\rm{abs}}$.
To see how each term contributes to the HHG yield, we remove each
term successively from our model. The results (envelope of HHG
spectra only) are shown in Figs.~\ref{fig4}(a) and \ref{fig4}(b) for
two pressures, 10 Torr and 50 Torr, respectively. At 10 Torr,
$\delta_{1}$, $\eta_{2}$, and $\delta_{\rm{h}}$ are small, their
effects  can be neglected, but the absorption term $\beta_{\rm{h}}$
is important, especially for low harmonics. [Note that $f_2$ is
large below 25 eV, see Fig.~\ref{fig4}(c)]. When the gas pressure is
increased to 50 Torr, the absolute values of $\delta_{1}$,
$\eta_{2}$, and $\delta_{\rm h}$ all increase by five times. Their
effects become  significant. In other words, with increasing gas
pressure, the change of coherence length cannot be attributed to one
single factor only.

For the absorption effect, we first calculate the absorption length
\emph{L}$_{\rm{abs}}$ for H15. They are about 2.4, 0.8, and 0.5 mm
at the gas pressures of 10, 30, and 50 Torr, respectively. According
to Eq.~(\ref{con-model}), only when \emph{L}$_{\rm{med}}>$
\emph{L}$_{\rm{abs}}$ the absorption effect would become important.
This has been verified in our calculation by comparing H15 at two
gas pressures without $\beta_{\rm h}$ in the model. The absorption
effect becomes significant for all harmonics in the plateau at high
pressure (50 Torr). It is important only for low harmonics at low
pressure (10 Torr).

We also show atomic scattering factors $f_{1}$ and $f_{2}$ in
Fig.~\ref{fig4}(c), they are  related to $\delta_{\rm h}$ and
$\beta_{\rm h}$ respectively~\cite{JinPRA2011}. We can see that the
change of harmonics with or without $\delta_{\rm h}$ (or $\beta_{\rm
h}$) follows the energy dependence of $f_{1}$ (or $f_{2}$). From
Fig.~\ref{fig4}, we conclude that there is not a single dominant
factor that determines the pressure dependence on the HHG observed
in Figs.~\ref{fig3}(a) and \ref{fig3}(b). Since these factors are
wavelength dependent, the effects vary with the harmonic order.

\subsubsection{Dispersion and absorption data from different sources}
\begin{figure}
\includegraphics[width=6.8cm]{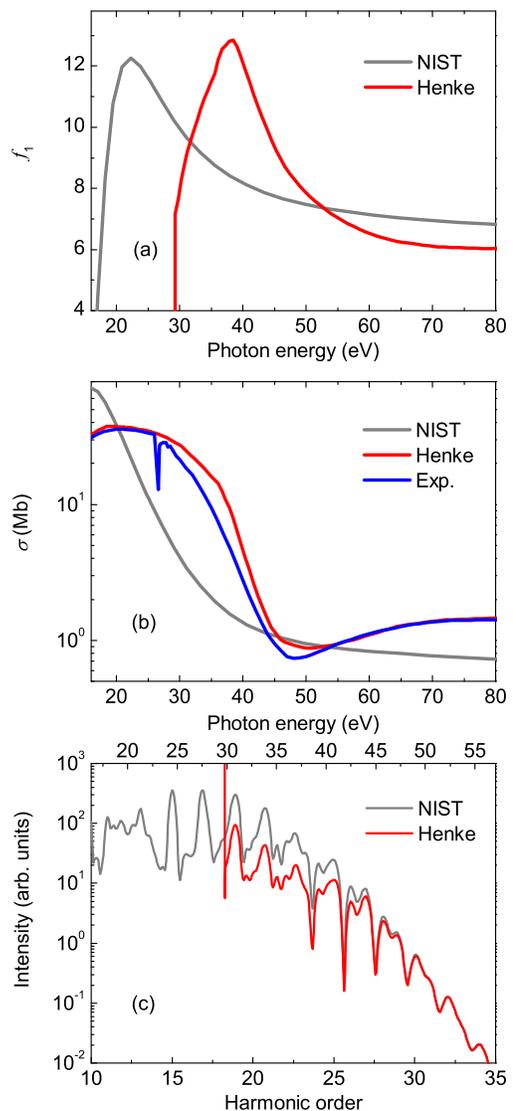} \vskip-0mm
\caption{(a) Comparison of atomic scattering factors $f_1$ from
Henke~\cite{Henke} and NIST~\cite{NIST} for Ar. (b) Comparison of
photoionization cross section between experiment~\cite{PI} and
theory from Henke~\cite{Henke} and NIST~\cite{NIST} by using the
formula $\sigma=2r_0\lambda f_2$. (c) Comparison of harmonic spectra
calculated by using different $f_1$, $f_2$ from \cite{Henke} and
\cite{NIST}, the laser peak intensity is 2.0$\times$10$^{14}$
W/cm$^2$, and gas pressure is 50 Torr.} \label{fig5}
\end{figure}

In Fig.~\ref{fig4}, we have used the atomic scattering factors
$f_{1}$ and $f_{2}$ from NIST~\cite{NIST} to calculate the
dispersion ($\delta_{\rm h}$) and absorption ($\beta_{\rm h}$)
effects in the propagation equation. An alternative set of atomic
scattering factors are given by Henke~\cite{Henke}. These two sets
of data are compared in Figs.~\ref{fig5}(a) and~\ref{fig5}(b). Since
$f_{2}$ is related to atomic photoionization cross section through
$\sigma=2r_0\lambda f_2$, where $r_0$ is the classical electron
radius, we can obtain theoretical $\sigma$ from these two sources to
compare with experimental photoionization cross sections tabulated
in \cite{PI}. We note that the NIST data underestimate the
absorption cross sections quite significantly. This would have the
effect of overestimating the simulated HHG yields. Using these two
sets of atomic scattering factors to simulate the HHG spectra, as
shown in Fig.~\ref{fig5}(c), there are significant differences
between the two calculations for the lower harmonics. Such
differences have not been addressed in the literature as far as we
know. However, the data set from Henke~\cite{Henke} did not extend
$f_{1}$ below 29 eV. In fact, it drops precipitously below about 40
eV. These atomic scattering factors were calculated originally for
x-ray energy region. They are perhaps not very reliable in the
photon energies considered here. The accuracy of scattering factors
in this energy region probably should be used with caution.

\subsection[]{Pressure dependence of HHG at different laser intensities and gas-jet lengths}
\begin{figure*}
\includegraphics[width=14.6cm]{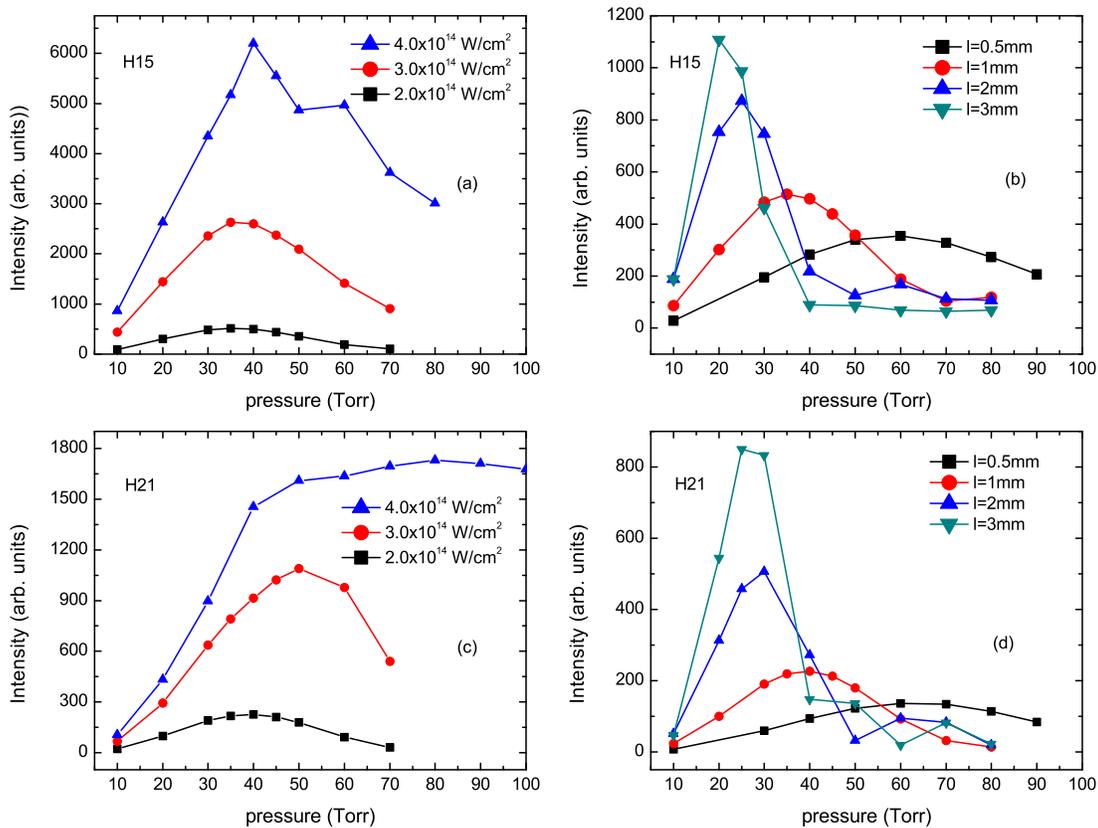} \vskip-1mm
\caption{The intensity of harmonics 15th and 21st as a function of
gas pressure for (a) and (c): different laser intensities, and (b)
and (d): different gas jet lengths. In (a) and (c) the gas jet is
1-mm long. In (b) and (d) the laser intensity at focus is
2.0$\times$10$^{14}$ W/cm$^2$. The other parameters are the same as
simulation-1 in Fig.~\ref{fig1}.} \label{fig6}
\end{figure*}

Figures~\ref{fig6}(a) and \ref{fig6}(c) compare the pressure
dependence of H15 and H21 at three different intensities in a gas
jet with the length of 1 mm. For each intensity, the pressure
dependence is similar to what we have seen in Fig.~\ref{fig3}. As
the intensity is increased, we note the highest harmonic yield
occurs at a higher pressure. Comparing the two harmonics, the
optimum harmonic yield also moves to a higher pressure for the
higher harmonic. For intensity of 4.0$\times$10$^{14}$ W/cm$^2$, H21
achieves optimal yield at a pressure close to 100 Torr. Because
higher intensity results in higher free electron density,
consequently, a higher gas pressure is needed to compensate for the
phase mismatch due to the free electrons. This means that if we want
to obtain high intensity for the high harmonics, we should use a
high laser intensity and a high gas pressure. This result is
consistent with the generation of water window harmonics using
mid-infrared lasers reported recently by Popmintchev \emph{et al.}
\cite{Tenico,Tenico2}, where gas pressure as high as a few
atmospheres were used, together with a higher laser intensity.

In Figs.~\ref{fig6}(b) and \ref{fig6}(d) we show the harmonic
intensities for H15 and H21 at several different lengths of the gas
jet (0.5, 1, 2, and 3 mm) using  laser intensity of
2$\times$10$^{14}$ W/cm$^2$ at focus (for each length, the center of
gas jet is always placed at 2 mm after laser focus). Below 20 Torr,
the harmonic intensity increases rapidly by increasing the
interaction length and gas pressure. When we further increase the
pressure to a higher value, on the contrary, the harmonic intensity
drops with the medium length. The optimal gas pressure increases
with decreasing interaction length, because the absorption length is
decreased with increasing gas pressure. The optimized harmonic
intensity with a long gas jet is higher than that of a short one.
Thus we can obtain a high harmonic intensity using a long gas medium
with low gas pressure. These results are very similar to those
observed by Tamaki {\it et al.} \cite{Midorikawa} using a 30-fs
laser pulse.

\subsection{Macroscopic HHG spectra of Ne: Theory vs experiment}

\begin{figure}
\includegraphics[width=8.0cm]{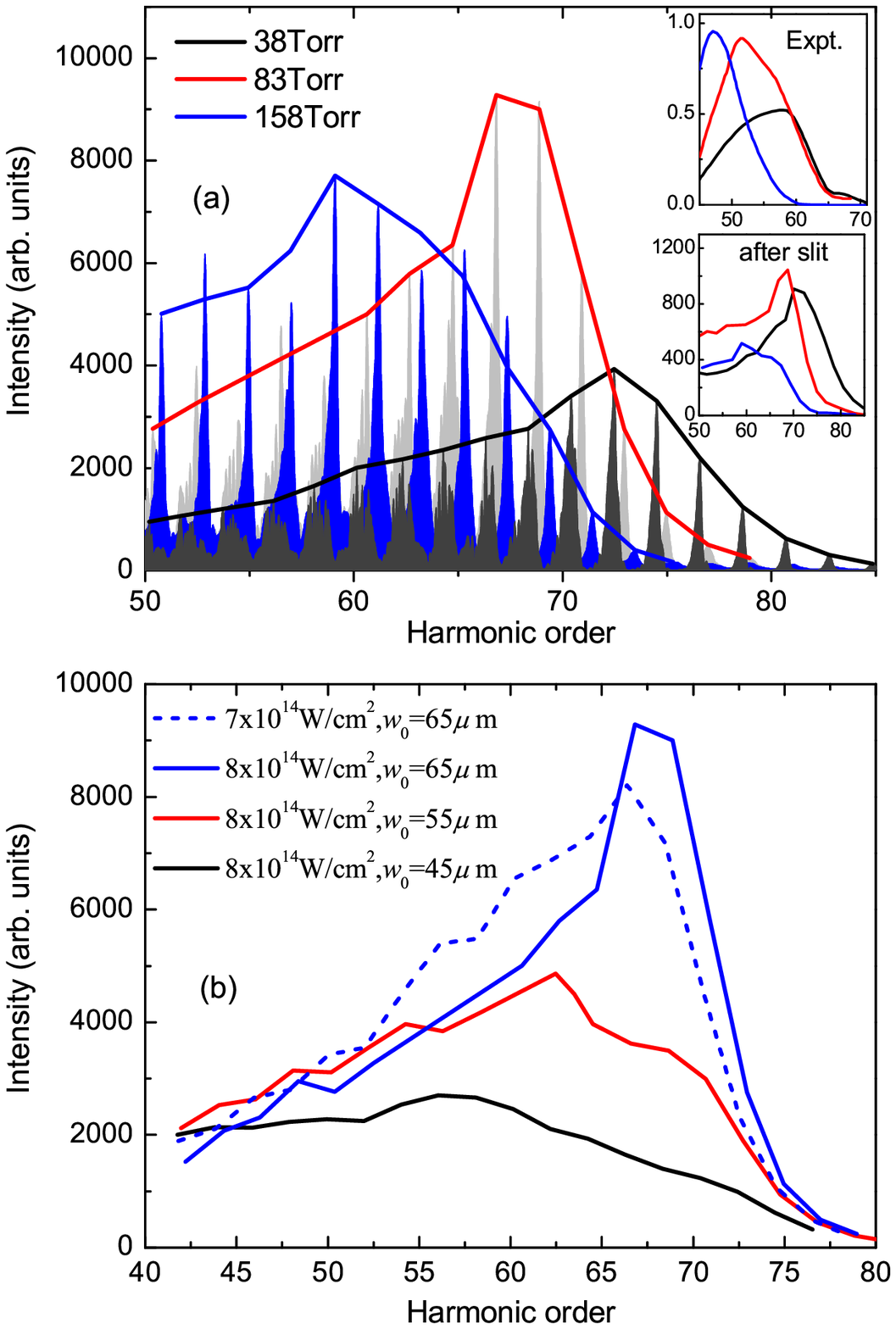} \vskip-1mm
\caption{(a) The simulation of total HHG spectra of Ne for three gas
pressures, 38, 83, and 158 Torr, generated by an 805-nm and 45-fs
laser field. The laser peak intensity at focus is 8$\times$10$^{14}$
W/cm$^2$. The first inset shows the envelope of experimental spectra
\cite{Dachraoui09}. The second inset shows the theoretical spectra
if the harmonics are taken with a slit of width of 500 $\mu$m placed
50 cm after the gas jet. (b) The influence of beam waist and laser
intensity on the simulated spectra for a pressure of 83 Torr. Other
parameters are given in the text.} \label{fig7}
\end{figure}

To test how the QRS model works for other atomic targets, we perform
a simulation of HHG spectra for Ne reported by Dachraoui {\it et
al.} \cite{Dachraoui09} recently. Figure~\ref{fig7}(a) shows the
comparison of HHG spectra between theory and experiment. The
experiment \cite{Dachraoui09} was performed using a laser with pulse
duration of 45 fs, central wavelength of 805 nm. The laser focusing
position was in the middle of the 3-mm long gas cell and the spot
size was estimated to be $\sim$110 $\mu$m (the beam waist at focus
was 55 $\mu$m). The peak vacuum intensity was
1.2($\pm$0.4)$\times$10$^{15}$ W$/$cm$^2$. Three different
pressures: 38, 83, and 158 Torr were used. In the simulation the
beam waist is 65 $\mu$m, the laser peak intensity is
8$\times$10$^{14}$ W$/$cm$^2$ (near the critical intensity of Ne,
which is $\sim$8.5$\times$10$^{14}$ W$/$cm$^2$), the other
parameters are the same as those given in the experiment. In
Fig.~\ref{fig7}(a) simulated spectra at the three pressures are
shown, together with the envelope of the calculated spectra. The
general trend of the spectra vs gas pressure agrees reasonably with
the experiment, see inset. Under present conditions the ionization
probability is very high ($\sim$75\% at the end of laser pulse with
peak intensity of 8$\times$10$^{14}$ W$/$cm$^2$ according to ADK
tunneling model \cite{ADK}). The fundamental field is strongly
reshaped during the propagation in the medium and has a strong
dependence on pressure. By increasing the gas pressure, the harmonic
cutoff position shifts to lower orders.

There are still large discrepancies between the experiment and the
simulation, especially the cutoff positions. The cutoff positions in
the theory are  about 10 harmonic orders higher than those in the
experiment. In our simulations, we find that many experimental
parameters would affect the shape and the cutoff of the spectra
significantly. In Fig.~\ref{fig7}(b) we show the influence of beam
waist and laser intensity on the spectra at a pressure of 83 Torr,
the other parameters are fixed as those in Fig.~\ref{fig7}(a). We
can see that the shape and cutoff position obtained under different
conditions are very different. The change of cutoff position for the
beam waist from 65 to 45 $\mu$m is about 10 harmonic orders. A
slight change of the laser intensity from 8 to
7$\times$10$^{14}$W$/$cm$^2$ would modify the shape of the HHG
spectra drastically. Since many factors mentioned above in the
experiment are not well determined, we have not been able to
simulate the observed spectra. We also comment that the experimental
HHG spectra depend critically on how they are measured
\cite{JinPRA2011}. For example, if a slit of width of 500 $\mu$m is
placed at 50 cm behind the gas jet, then only  harmonics emitted
close to the axis will be measured. The second inset in
Fig.~\ref{fig7}(a) shows how the envelopes of the harmonic spectra
at the three pressures are modified. For higher pressure, there are
more free electrons generated and thus the fundamental laser is more
defocused as the pressure is increased. The inset shows that the
relative yield at the higher pressure is reduced, indicating that
HHG generated at higher pressure is more divergent.

\section{Summary and outlook}

In summary, we have used the recently developed quantitative
rescattering (QRS) theory for the generation of high-order harmonics
of individual atoms with the inclusion of the propagation of the
laser and the harmonics in the gas medium to obtain macroscopic HHG
spectra that can be compared directly with experimental
observations. We have studied the effect of the gas pressure on the
yields of the harmonics. The gas pressure affects harmonics through
frequency-dependent properties of the medium such as dispersion,
absorption, nonlinear Kerr effect, and plasma density, as well as
the phase-matching conditions. In the meanwhile, the fundamental
laser field is modified as it propagates through the medium. The
modification is especially severe when the incident laser reaches
the intensity near or above the critical intensity for a given
target. From the simulations, we have found that HHG spectra are
very sensitive to the laser focusing conditions. By starting with
the QRS theory instead of the SFA for the single-atom response, we
have been able to obtain HHG spectra that are much closer to
experimental observations, but we are still unable to obtain HHG
spectra that are in perfect agreement with the data. This is in
contrast with our recent positive results
\cite{JinPRA2011,JinJPB,Jin-CO2} where HHG of atoms and molecules
were generated at lower intensities. We have attributed the lack of
good agreement to the strong dependence of the HHG spectra on the
laser parameters when experiments are carried out at high
intensities and that these parameters are difficult to specify
accurately in a given experiment. In the future, it is desirable
that experiments report HHG spectra taken at various focusing
conditions and gas pressures in order to establish if our current
theory of HHG is under solid ground in the high-intensity and
high-gas-pressure regime. This regime is important in order to
generate high-energy harmonics near and above the water window using
mid-infrared lasers, as well as the possible generation of few-cycle
single attosecond pulses in the soft X-ray region
\cite{Tenico,Tenico2,light-source}.

\section{Acknowledgments}
This work was supported in part by Chemical Sciences, Geosciences
and Biosciences Division, Office of Basic Energy Sciences, Office of
Science, U.S. Department of Energy. G.-L.W was also supported by the
National Natural Science Foundation of China under Grant Nos.
11064013 and 11044007, the Specialized Research Fund for the
Doctoral Program of Higher Education of China under Grant No.
20096203110001 and the Foundation of Northwest Normal University
under Grant No. NWNU-KJCXGC-03-70.


\begin{thebibliography}{}

\bibitem{Tenico}T. Popmintchev, M.-C. Chen, A. Bahabad, M. Gerrity,
P. Sidorenko, O. Cohen, I. P. Christov, M. M. Murnane, and H. C.
Kapteyn, Proc. Natl. Acad. Sci. USA \textbf{106}, 10516 (2009).

\bibitem{Tenico2} T. Popmintchev, M.-C. Chen, P. Arpin, M. M. Murnane, H. C.
Kapteyn, Nature Photon. \textbf{4}, 822 (2010).

\bibitem{light-source}M.-C. Chen, P. Arpin, T. Popmintchev, M. Gerrity, B. Zhang, M.
Seaberg, D. Popmintchev, M. M. Murnane, and H. C. Kapteyn, Phys. Rev.
Lett. \textbf{105}, 173901 (2010).

\bibitem{Atto1}M. F. Kling and M. J. J. Vrakking, Annu. Rev. Phys. Chem.
\textbf{59}, 463 (2008).

\bibitem{Atto2}F. Krausz and M. Ivanov, Rev. Mod. Phys. \textbf{81}, 163
(2009).

\bibitem{JinJPB}C. Jin, H. J. W\"{o}rner, V. Tosa, A. T. Le, J. B. Bertrand,
R. R. Lucchese, P. B. Corkum, D. M. Villeneuve, and C. D. Lin, J.
Phys. B \textbf{44}, 095601 (2011).

\bibitem{JinPRA2011}C. Jin, A. T. Le, and C. D. Lin, Phys. Rev. A
\textbf{83},
023411 (2011).

\bibitem{Jin09}C. Jin, A. T. Le and C. D. Lin, Phys. Rev. A \textbf{79},
053413 (2009).

\bibitem{Jin-CO2}C. Jin, A. T. Le and C. D. Lin, Phys. Rev. A \textbf{83}, 053409
(2011).

\bibitem{qrs1}T. Morishita, A. T. Le, Z. Chen, and C. D. Lin, Phys. Rev. Lett.
\textbf{100}, 013903 (2008).

\bibitem{qrs2}A. T. Le, R. R. Lucchese, S. Tonzani, T. Morishita, and C. D.
Lin, Phys. Rev. A \textbf{80}, 013401 (2009).

\bibitem{qrs3}C. D. Lin, A. T. Le, Z. Chen, T. Morishita, and R. R. Lucchese,
J. Phys. B \textbf{43}, 122001 (2010).


\bibitem{SFA}M. Lewenstein, Ph. Balcou, M. Yu. Ivanov, A. L'Huillier,
and P. B. Corkum, Phys. Rev. A \textbf{49}, 2117 (1994).

\bibitem{Altucci05}C. Altucci, R. Velotta, J. P. Marangos, E. Heesel, E. Springate, M. Pascolini, L.
Poletto, P. Villoresi, C. Vozzi, G. Sansone, M. Anscombe, J-P. Caumes, S. Stagira, and M. Nisoli, Phys. Rev. A \textbf{71}, 013409 (2005).

\bibitem{Japan}S. Minemoto, T. Umegaki, Y. Oguchi, T. Morishita, A. T. Le,
S. Watanabe, and H. Sakai, Phys. Rev. A \textbf{78}, 061402(R)
(2008).

\bibitem{Dachraoui09}H. Dachraoui, T. Auguste, A. Helmstedt, P. Bartz, M. Michelswirth, N. Mueller, W. Pfeiffer, P
Salieres, and U. Heinzmann, J. Phys. B \textbf{42}, 175402 (2009).

\bibitem{ADK}M. V. Ammosov, N. B. Delone, and V. P. Krainov, Zh. Eksp.
Teor. Fiz. \textbf{91}, 2008 (1986) [Sov. Phys. JETP \textbf{64},
1191 (1986)].

\bibitem{Tong}X. M. Tong and C. D. Lin, J. Phys. B \textbf{38}, 2593 (2005).

\bibitem{Muller}H. G. Muller, Phys. Rev. A \textbf{60}, 1341 (1999).

\bibitem{Bessel1}H-C. Bandulet, D. Comtois, A. D. Shiner, C. Trallero-Herrero,
N. Kajumba, T. Ozaki, P. B. Corkum, D. M. Villeneuve, J-C. Kieffer,
and F. L\'{e}gar\'{e}, J. Phys. B \textbf{41}, 245602 (2008).

\bibitem{Bessel2}M. Nisoli, E. Priori, G. Sansone, S. Stagira, G. Cerullo,S. De
Silvestri, C. Altucci, R. Bruzzese, C. de Lisio, P. Villoresi, L.
Poletto, M. Pascolini, and G. Tondello, Phys. Rev. Lett.
\textbf{88}, 033902 (2002).

\bibitem{Bessel3}C. Altucci, R. Bruzzese, C. de Lisio, M. Nisoli, E. Priori, S. Stagira, M. Pascolini, L. Poletto, P. Villoresi, V. Tosa, and K. Midorikawa, Phys. Rev. A \textbf{68}, 033806 (2003).

\bibitem{Hans-prl-09}H. J. W\"{o}rner, H. Niikura, J. B. Bertrand, P. B. Corkum, and D. M. Villeneuve, Phys. Rev. Lett. \textbf{102}, 103901 (2009).

\bibitem{Colosimo-NatPhys08}P. Colosimo {\it et al.}, Nat. Phys. \textbf{4},
386 (2008).

\bibitem{stanford}J. P. Farrell, L. S. Spector, B. K. McFarland, P. H. Bucksbaum, M. G\"{u}hr, M. B. Gaarde, and K. J.
Schafer, Phys. Rev. A \textbf{83}, 023420 (2011).

\bibitem{Jin-Atto} C. Jin, A. T. Le, C. A. Trallero-Herrero, and C. D. Lin, e-print arXiv:1107.3609.

\bibitem{APL}Y. H. Zheng, Z. N. Zeng, H. Xiong, R. X. Li, Z. Z. Xu, Y. Peng, X. Yang, and H. P. Zeng, Appl. Phys. Lett. \textbf{95}, 141102
(2009).

\bibitem{Krausz2000}T. Brabec and F. Krausz, Rev. Mod. Phys. \textbf{72}, 545 (2000).

\bibitem{Schnurer99}M. Schn\"{u}rer, Z. Cheng, M. Hentschel, G. Tempea, P. K\'{a}lm\'{a}n, T. Brabec, and F. Krausz. Phys. Rev. Lett. \textbf{83}, 722 (1999).

\bibitem{Schnurer2000}M. Schn\"{u}rer, Z. Cheng, M. Hentschel, F. Krausz, T. Wilhein, D. Hambach, G. Schmahl, M. Drescher, Y.
Lim, U. Heinzmann, Appl. Phys. B \textbf{70} [Suppl.], S227 (2000).

\bibitem{Altucci96}C. Altucci, T. Starczewski, E. Mevel, and C.-G. Wahlstr\"{o}m, J. Opt. Soc. Am. B \textbf{13}, 148
(1996).

\bibitem{Rundquist98}A. Rundquist, C. G. Durfee III, Z. Chang, C. Herne, S. Backus, M. M. Murnane, and H. C. Kapteyn, Science \textbf{280}, 1412 (1998).

\bibitem{Durfee99}C. G. Durfee, A. R. Rundquist, S. Backus, C. Herne, M. M. Murnane, and H. C. Kapteyn, Phys. Rev. Lett. \textbf{83} 2187 (1999).

\bibitem{Xu04} X. H. Xie, Z. N. Zeng, R. X. Li, S. Chen, H. H. Lu, D. J. Yin, and Z. Z. Xu, Science in China Ser. G Physics, Mechanics \&
Astronom \textbf{47}, 492 (2004).

\bibitem{Constant}E. Constant, D. Garzella, P. Breger, E. M\'{e}vel, Ch. Dorrer, C. Le Blanc, F. Salin, and P. Agostini, Phys. Rev. Lett. \textbf{82}, 1668 (1999).


\bibitem{NIST}C. T. Chantler, K. Olsen, R. A. Dragoset, J. Chang, A.
R. Kishore, S. A. Kotochigova, and D. S. Zucker, \emph{X-ray Form
Factor, Attenuation and Scattering Tables} (version 2.1) (National
Institute of Standards and Technology, Gaithersburg, MD, 2005)
[http://physics.nist.gov/ffast].


\bibitem{Henke}B. L. Henke, E. M. Gullikson, and J. C. Davis, At. Data Nucl.
Data Tables \textbf{54}, 181 (1993), available
from:\url{http://henke.lbl.gov/optical_constants/}.

\bibitem{PI}J. A. R. Samson and W. C. Stolte, J. Electron Spectrosc. Relat.
Phenom. \textbf{123}, 265 (2002).


\bibitem{Midorikawa}Y. Tamaki, J. Itatani, M. Obara, and K. Midorikawa,
Phys. Rev. A \textbf{62}, 063802 (2000).



\end{thebibliography}
\end{document}